# Multi-instrumental observations of the 2014 Ursid meteor outburst

Manuel Moreno-Ibáñez[1,2], Josep Ma. Trigo-Rodríguez[1], José María Madiedo[3,4], Jérémie Vaubaillon[5], Iwan P. Williams[6], Maria Gritsevich[7,8,9], Lorenzo G. Morillas[10], Estefanía Blanch[11], Pep Pujols[12], François Colas[5], and Philippe Dupouy[13]

[1]*Institute of Space Sciences (IEEC-CSIC), Meteorites, Minor Bodies and Planetary Science Group, Campus UAB, Carrer de Can Magrans, s/n E-08193 Cerdanyola del Vallés, Barcelona, Spain*
*mmoreno@ice.csic.es*; *trigo@ice.csic.es*
[2]*Finnish Geospatial Research Institute, Geodeetinrinne 2, FI-02431 Masala, Finland.*
[3]*Facultad de Ciencias Experimentales, Universidad de Huelva, 21071 Huelva, Spain.*
[4]*Departamento de Física Atómica, Facultad de Física, Universidad de Sevilla, Molecular y Nuclear, 41012 Sevilla, Spain.*
[5]*Institut de mécanique céleste et de calcul des éphémérides - Observatoire de Paris PSL, 77 avenue Denfert-Rochereau, 75014 Paris, France.*
[6] *School of Physics and Astronomy, Queen Mary University of London, E1 4NS.*
[7]*Department of Physics, University of Helsinki, Gustaf Hällströmin katu 2a, P.O. Box 64, FI-00014 Helsinki, Finland.*
[8]*Department of Computational Physics, Dorodnicyn Computing Centre, Federal Research Center "Computer Science and Control" of the Russian Academy of Sciences, Vavilova str. 40, Moscow, 119333 Russia.*
[9]*Institute of Physics and Technology, Ural Federal University,Mira str. 19. 620002 Ekaterinburg, Russia.*
[10] *IES Cañada de las Fuentes, Calle Carretera de Huesa 25, E-23480, Quesada, Jaén, Spain.*
[11]*Observatori de l'Ebre (OE, CSIC - Universitat Ramon Llull), Horta Alta, 38, 43520 Roquetes, Tarragona, Spain.*
[12]*Agrupació Astronómica d'Osona (AAO), Carrer Pare Xifré 3, 3er. 1a. 08500 Vic, Barcelona, Spain.*
[13]*Dax Observatoire, Rue Pascal Lafitte, 40100 Dax, France.*

## Abstract

The Ursid meteor shower is an annual shower that usually shows little activity. However, its Zenith Hourly Rate sometimes increases, usually either when its parent comet, 8P/Tuttle, is close to its perihelion or its aphelion. Outbursts when the comet is away from perihelion are not common and outburst when the comet is close to aphelion are extremely rare. The most likely explanation offered to date is based on the orbital mean motion resonances. The study of the aphelion outburst of December 2000 provided a means of testing that hypothesis. A new aphelion outburst was predicted for December 2014. The Spanish Meteor Network in collaboration with the French Fireball Recovery and InterPlanetary Observation Network set up a campaign to monitor this outburst and eventually retrieve orbital data that expands and confirms previous preliminary results and predictions. Despite unfavourable weather conditions over the South of Europe over the relevant time period precise trajectories from multi-station meteor data recorded over Spain were obtained, as well as orbital and radiant information for four Ursid meteors. The membership of these four meteors to the expected dust trails that were to provoke the outburst is discussed, and we characterize the origin of the outburst in the dust trail produced by the comet in the year 1392 A.D.

## 1.- INTRODUCTION

The first identification of a possible meteor shower with a radiant point in the constellation of Ursa minor was by Denning (1912), occurring around the Winter Solstice. However, very few observations were subsequently made. Hoffmeister recorded its activity in 1914, but its existence was not confirmed until 1945 (Becvár 1946). There are two reasons for the lack of records of the Ursids. First the Ursids have very low activity levels in most years, its Zenith Hourly Rate (the number of meteors that would be observed under good observing conditions in one hour if the radiant was at the zenith) is usually ZHR<10 (Jenniskens 1994). Second, the weather can be bad in mid December and many observers choose to only observe the more predictable Geminids. Ceplecha (1951) showed that the annual Ursid shower was related to comet 8P/Tuttle, a Jupiter family comet with a period of 13.6 years (JPL-HORIZONS, ssd.jpl.nasa.gov) and a perihelion distance slightly greater than 1 AU. This means that the comet is usually brightest (close to perihelion) and close to the Earth at roughly the same time. Hence it has been observed at all perihelion passages since discovery apart from 1953, when observing conditions were poor throughout. These perihelion passages were in 1858, 1872, 1885, 1899, 1913, 1926, 1940, 1953, 1967, 1980, 1994 and 2008. As it can take a considerable time after being ejected from the nucleus for meteoroids to disperse away from the nucleus locality, an enhancement is generally to be expected in stream activity at the time when the comet is close to perihelion and new meteoroids are injected (see Williams, Johnson & Fox 1986 for an early discussion and mathematical formulation of this). An enhancement at some such times has also been observed in the Ursids, where the ZHR reaches around 3 times the normal rate (e.g. 1900, 1914, 1953, 1981 and 1996, data taken from www.meteorshowersonline.com). However, in the case of the Ursids, several perihelion passage years do not show any significant enhancement and so this is unlikely to be the explanation. Instead, Jenniskens et al. (2007) suggested that they are caused by cometary material released at very old comet perihelion passages between AD 300-1400 and appear as a wide and stretched stream (which we will call and refer to as a “filament”).

What is remarkable about the Ursids is that a much sharper increase in the ZHR to well over 100 was observed in 1795, 1945 and 1986 (Jenniskens 1994) and 2000 (Jenniskens et al. 2002), years when the parent comet is near aphelion at a heliocentric distance of over 10 AU, so that it will not be outgassing and inserting new meteoroids into the stream at this point.

Two body mean motion resonances can play an important role in determining the behaviour of meteor showers as was first pointed out by Asher, Bailey and Emel’yanenko (1999) who showed the importance of Jovian resonances in relation to the Leonids storms. Other resonance effects have also been found in streams, for example the two body Saturnian resonances (Sekhar and Asher 2013), two body Uranian resonances (Williams 1997) and three body resonances involving both Jupiter and Saturn (Sekhar, Asher and Vaubaillon 2016). Jenniskens et al. (2002), pointed out that this effect can play an important role in the Ursids. The comet is trapped in a 15:13 resonance with Jupiter, and so its period is fixed at about 13.6 years. The orbits of the meteoroids initially released near the comet perihelion slowly evolve due to the radiation pressure and get trapped in a 7:6 mean motion resonance with Jupiter and so effectively now also have a roughly fix period determined by this resonance. This means that the ratio between the meteoroid period and the comet period is also roughly fixed and is (13x7/15x6) = 1.011. Hence, in 45 or 46 orbits (around 620 years), the comet and meteoroids will be exactly out of phase (one at its perihelion, other at its aphelion).

A very interesting coincidence was both theoretical and computationally predicted to occur in 2014. Meteoroids in the filament and trapped in the resonance were both predicted to reach the Earth in December 2014. In order to properly record this exceptional event, the SPanish Meteor Network (SPMN) set up a campaign with the collaboration of the recently established French Fireball Recovery and InterPlanetary Observation Network (FRIPON) to monitor the 2014 Ursid meteor shower. The use of several SPMN high sensitivity CCD video devices allowed us to record Ursid meteors with good spatial resolution. This paper describes these observations and their results, focusing mainly on the aphelion outburst.

As might be expected in December, bad weather hampered the continuous observations carried out at southern SPMN stations. However, three Spanish north-east stations located in Catalonia as well as south stations in France succeeded in detecting these meteors during the night of the 22$^{nd}$ to 23$^{rd}$ of December 2014.

## 2.- OUTBURST PREDICTIONS

A simulation of the encounter with possible dust trails was carried out using the software developed by Vaubaillon, Colas and Jorda (2005a, b) and used for the planning of the 2014 SPMN-FRIPON campaign. The simulation involved computing the evolution of possible dust trails released from the comet 8P/Tuttle at 21 perihelion passages between AD 351 and AD 2008. It transpired that only meteoroids released in year 1392 evolved in a way that would encounter the Earth in 2014 (exactly in line with the calculation given in the introduction of requiring about 620 years for the meteoroids to evolve). This encounter was predicted to be on December $23^{rd}$ at 00:46h (UT), at a solar longitude of $\lambda_0$=270.743º. The reliability of the predictions can be evaluated using a Confidence Index defined by Vaubaillon (2016). For the 1392 meteoroid stream this index is SYO0/1CE0.00, indicating that the simulated particles belong to a single trail, crossing the Earth in a single year, and no close encounters with other planets that would have altered the orbit (e.g. Dmitriev, Lupovka and Gritsevich 2015).

The nodes of its orbit and the Earth trajectory are plotted in Fig. 1.

Jenniskens (2006) claimed that the most likely trails to cause the previous Ursid aphelion outbursts mentioned in the introduction are those released from the comet in 1392 and 1405. Jenniskens (2006) predicted that the 2014 Ursid outburst would be caused by a dust trail detached from the comet in 1405. This encounter was expected for December $22^{nd}$ 2014 at 23:38h (UT) at a solar longitude of $\lambda_0$=270.838º. Jenniskens (2006) also predicted that the Ursid filament would encounter the Earth at 17:05h (UT) on December $22^{nd}$, with $\lambda_0$=270.56º.

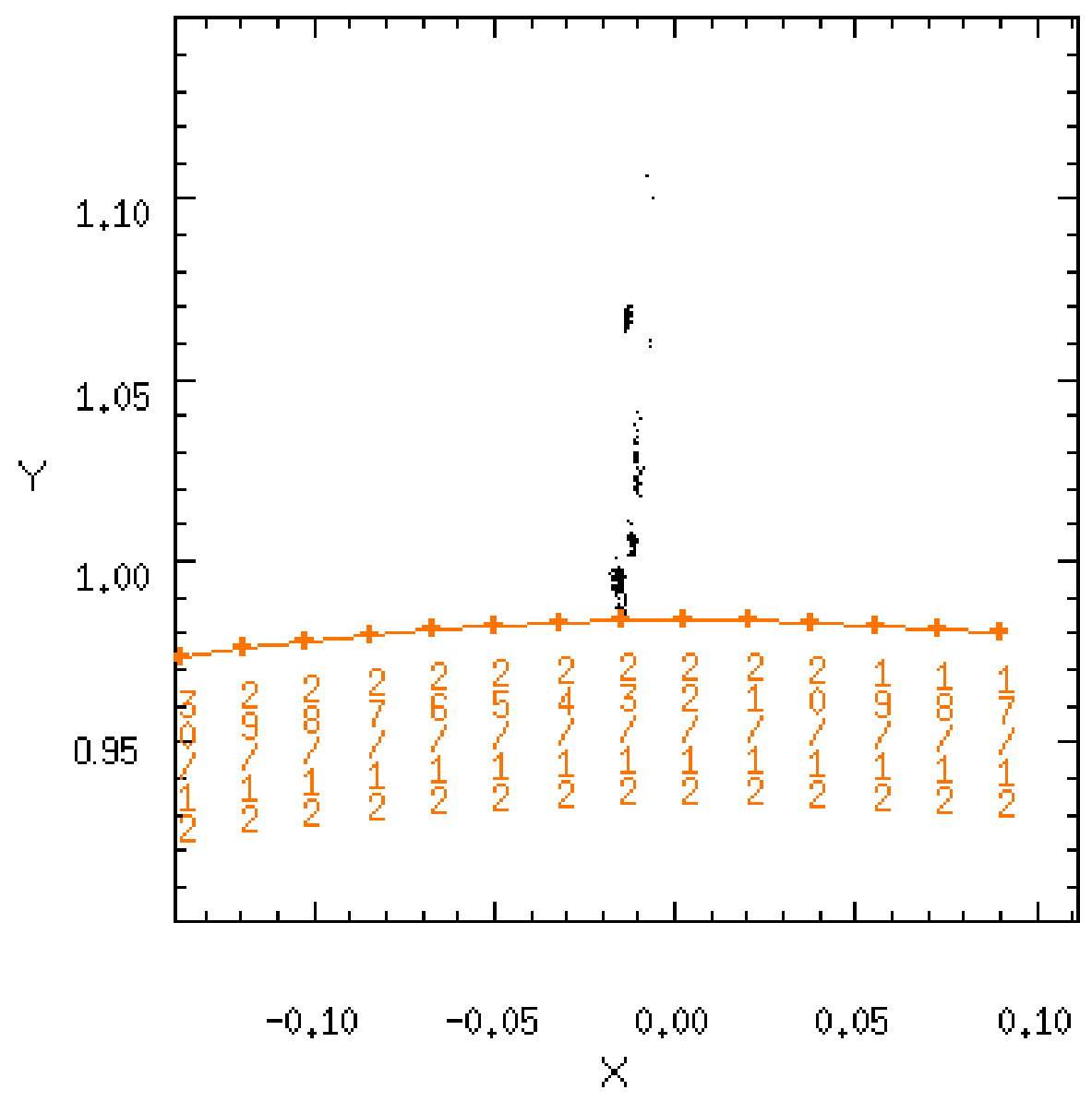

**Figure 1**. Nodes of the dust trails encountering the Earth in 2014. The trajectory of the Earth is also plotted. X and Y values indicate the coordinates on the ecliptic plane.

## 3.- INSTRUMENTATION AND DATA REDUCTION ANALYSIS

Several stations were involved in monitoring the expected Ursid meteor outburst in December 2014. The SPMN facilities use an array of low-light CCD video cameras (Water Co. models 920H and 902H Ultimate). These stations are equipped with the UFO Capture software that allows autonomous continuous monitoring of the sky. The video uses a 1/2" Sony interline transfer CCD image sensor with their minimum lux rating ranging from 0.01 to 0.0001 lx at f1.4. Their focal length ranges from 6 to 25 mm and the field of view covered by each device ranges from 90 to 120 degrees. The video cameras employ aspherical fast lenses with focal lengths ranging 4-12 mm, and focal ratios between 0.8-1.2. The detections are registered in a cluster of computers synchronize according to GPS devices, providing meteor recording with an accuracy of 0.1 seconds (Madiedo and Trigo-Rodríguez 2008; Madiedo et al. 2010). This system allows coverage of several regions of the sky with each camera, obtain point-like star images and detect meteors showing an apparent magnitude of +3±1. The video system used consists of a PAL standard with a resolution of 720 x 576 pixels.

The reduction pipeline has been described in Trigo-Rodríguez et al. (2004). The reduction software generates a composite image of the complete meteor atmospheric trajectory where the background stars can be identified. Then, a manual astrometric analysis is undertaken in order to retrieve the pixel position of the meteor trail and the surrounding stars that serve as calibrators. The NETWORK software was used in this step to derive the trajectory, velocity and radiant data (Trigo-Rodríguez et al. 2004), simply followed the methodology described in (Ceplecha 1987). The AMALTHEA software, developed by Madiedo, is used to derive computationally the orbital parameters (Trigo-Rodríguez, Madiedo and Williams. 2009; Trigo-Rodríguez et al. 2009; Madiedo, Trigo-Rodríguez and Lyytinen 2011).

A forward-scatter radio system operating at a frequency of 143.05 MHz was included in the observation campaign. This radio station is located in Jaén (Spain), and employs an 8 dBi six-element Yagi antenna and a Yaesu FT817 ND radio receiver, listening to the Grand Réseau Adapté à la Veille Spatialle (GRAVES) radar located in Dijon, France (http: //www.onera.fr/dcps/graves).

The FRIPON stations involved in the campaign are listed in Table 1. They have DMK 23G445 all sky cameras (from Imaging Source GmbH). This model contains a Sony Chip ICXX445 (1348 x 976 pixels, 5 x 4 mm). These cameras allow exposures of 30 ms and are equipped with a 1/3" sensor. A Focusave lens of 1.25 mm and f/d=2.0 has been chosen to fulfill the focal length requirements (from 1 to 1.5 mm) of the stations (Colas et al. 2014). The information registered is transmitted to the store disks through a GigE Vision protocol, which also allows the use of PoE protocol to deliver energy and power to the camera.

**4.- OBSERVATIONS**

Many observers reported an Ursid outburst in 2014. Gajdoš, Tóth and Kornoš (2015) registered 19 Ursids between 21h20m UT, Dec. 22 and 05h35m UT Dec. 23. These were single station detections using the AMOS all-sky camera and so they could not derive the orbital data for these meteors. The NASA's Camera for All-sky Meteor Surveillance (CAMS) project in California, detected 20 meteors during the main activity time (01h32m UT Dec.23 to 04h00m Dec. 23) and 15 more after this time (Jenniskens 2014) (Dec. 23) at $\lambda_0$=270.85º. The Canadian Meteor Orbit Radar (CMOR) detected up to 85 meteors between 23h15m UT (Dec. 22) and 0h45m UT (Dec. 23) (Brown 2014). There are also radio meteor detections (forward-scatter technique) by Yrjöla (Kuusankoski, Finland) reporting high Ursid meteor activity in this period (Jenniskens 2014).

The SPMN - FRIPON campaign optically detected 29 Ursid video meteors recorded at the SPMN stations listed in Table 1. They had a population index of 1.8±0.6 (Table 2), similar to the 1.7 found by Molau (2014). The mean Ursid ZHR was around 19, in agreement with the ZHR=10 prediction of Jenniskens (2006); but peaking with a ZHR=45 ±19 which is also close to the value reported in Molau (2014).

In Table 3 the number of Ursids, Coma Berenicids and sporadic meteors recorded by the three cameras covering the night sky from Folgueroles, the SPMN station with the darkest skies available that night, are compiled. Two significant increases in the Ursid hourly rate occurred around 23h45 and 04h45 UT. A forward-scatter technique operating at a radio station in Jaén (Andalusia, Spain) was used to detect the Ursid meteor activity. Despite the fact that the forward-scatter technique is unable to distinguish the source of each detection, the observations indicate significant activity between 00.00h and 01.00h on December 23$^{rd}$. In Fig. 2 the normalized echoes (over the mean number of echoes detected every hour) detected between 12h UT (Dec. 21) and 12h UT (Dec. 24) are plotted. The results prove an increasing activity that night starting close to the predicted time.

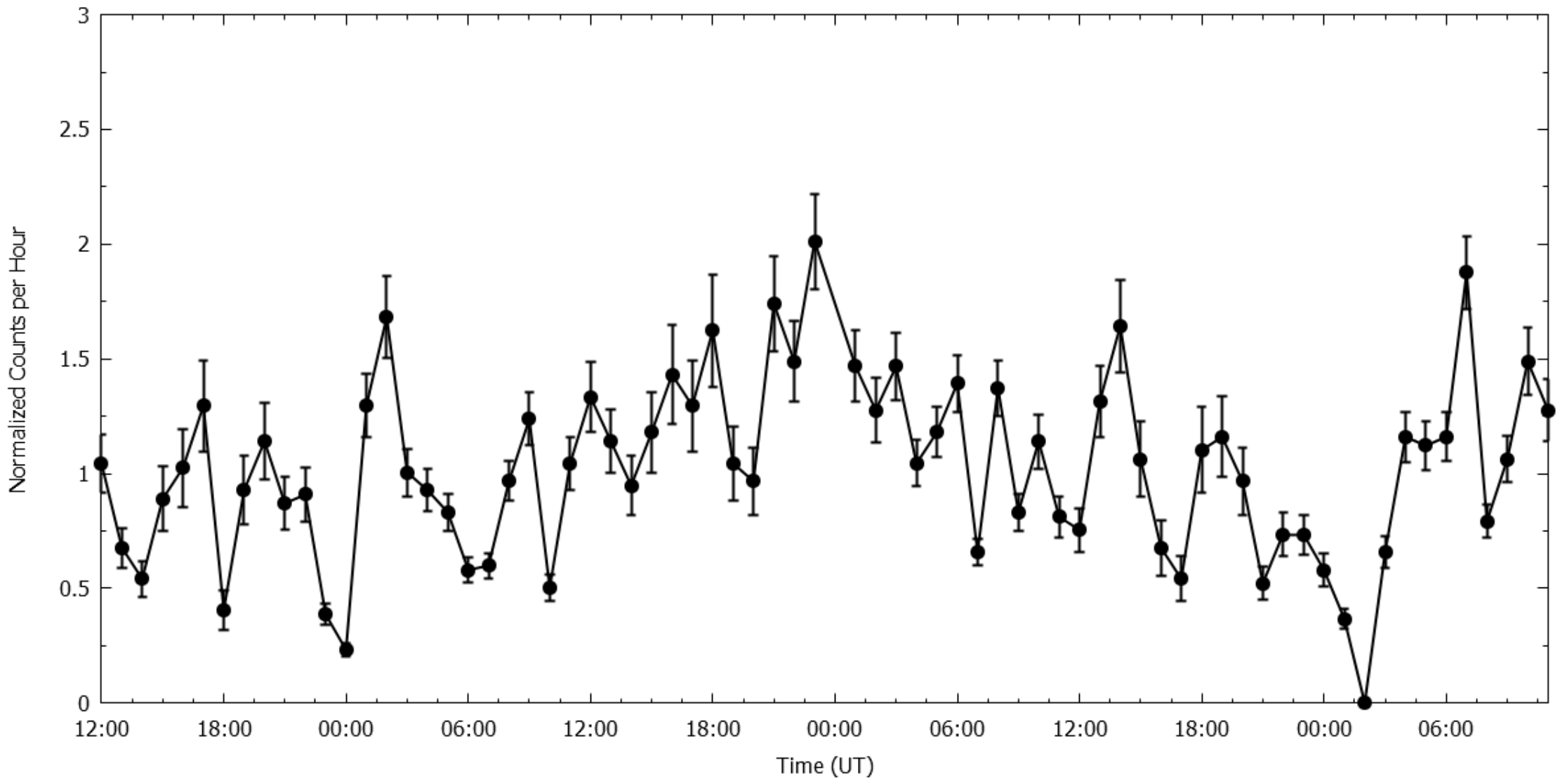

**Figure 2.** Radio meteors counts normalized per hour (over the mean number of counts each hour) detected at Jaén forward-scatter station, from December 21nd at 12h (UT) to December 24th at 12h (UT). An increase activity was observed in the night of 22nd to 23rd of December.

| Network | Station (Province) | Longitude | Latitude (N) | Alt. (m) | Imaging system |
|---|---|---|---|---|---|
| SPMN | Montsec (Lleida) | 00º 43´ 46" E | 42º 03´ 05" | 1570 | AS |
| SPMN | Montseny (Girona) | 02º 31´ 14" E | 41º 43´ 17" | 300 | WFV |
| SPMN | Folgueroles (Barcelona) | 02º 19´ 33" E | 41º 56´ 31" | 580 | WFV |
| SPMN | Ebre Observatory | 00º 29´ 44” E | 40º 49´16 “ | 50 | WFV |
| FRIPON | Pic du Midi | 00º 08´ 34" E | 42º 56´ 11" | 2876 | AS |
| FRIPON | Dax Observatory | 01º 01´ 49.8" W | 43º 41´ 36.4" | 470 | AS |

**Table 1.** Details of the SPMN and FRIPON stations involved in the Ursid aphelion outburst campaign. Acronyms for the different imaging systems are: AS (low-scan-rate CCD all-sky camera), and WFV (Wide field video cameras).

| **Magnitude** | **-5** | **-4** | **-3** | **-2** | **-1** | **0** | **+1** | **+2** | **+3** | **N** | **r** |
|---|---|---|---|---|---|---|---|---|---|---|---|
| Number | 1 | 2 | 1 | 1 | 2 | 4 | 6 | 8 | 5 | 30 | 1.8±0.6 |

**Table 2.** Magnitude distribution of all Ursids imaged by SPMN video cameras during the night of December 22nd-23rd, 2014.

| Time interval (UT) | Mean Solar longitude (º) | Ursids | Coma Berenicids | Sporadics |
|---|---|---|---|---|
| **18h15-19h15** | **271.64** | **1** | **0** | **1** |
| **19h15-20h15** | **271.68** | **3** | **0** | **0** |
| **20h15-21h15** | **271.73** | **3** | **0** | **0** |
| **21h15-22h15** | **271.77** | **1** | **0** | **0** |
| **22h15-23h15** | **271.81** | **2** | **1** | **1** |
| **23h15-00h15** | **271.85** | **6** | **0** | **3** |
| **00h15-01h15** | **271.89** | **1** | **3** | **1** |
| **01h15-02h15** | **271.93** | **1** | **0** | **1** |
| **02h15-03h15** | **271.97** | **1** | **2** | **3** |
| **03h15-04h15** | **272.02** | **2** | **2** | **10** |
| **04h15-05h15** | **272.07** | **6** | **3** | **8** |
| **05h15-06h15** | **272.11** | **2** | **3** | **3** |

**Table 3.** Ursids, Coma Berenicids and sporadic meteors recorded from the Folgueroles SPMN video station, given every hour during the night of 22nd-23rd December 2014. The meteor limiting magnitude recorded by the cameras was +4. The mean hourly rate was ZHR: 19 ±3, with a peak at 23h45m UTC of ZHR: 45 ±19.

## 5. RESULTS: TRAJECTORY, RADIANT AND ORBITAL DATA

As mentioned in the introduction, bad weather conditions over the Iberian Peninsula limited the observations to the northeast region of Spain and southeast of France. Despite this, the SPMN network successfully recorded 11 Ursid meteors at multiple stations. It had been envisaged that collaboration between the SPMN and FRIPON would provide multi-station observations of any recorded Ursid meteors. Unfortunately, most of the meteors were too faint for the FRIPON network to detect and the distance between the networks was too great so that the same meteors could not be detected by both networks.

The trajectory, radiant and orbital elements were derived following the intersection of planes methodology developed by Ceplecha (1987). The reliability of the derived results depends on the angle, Q, between the two planes containing the two observing stations and the trajectory in which the meteor is apparently moving. If Q is small, triangulation measurements become unreliable as the triangle has degenerated into a near straight line. The orbital results we report were derived from observations where Q was higher than 20º. As a consequence of the bad weather already mentioned, which made observing difficult or impossible from some stations, most of the active stations tended to be aligned in a rough North-South direction. Since the radiant point of the Ursids is close to the Pole Star, there was an inevitable preponderance of meteors moving in a roughly North-South direction. These two facts imply that for many observed meteors the angle Q was less than 20º and so the determination of trajectories was unreliable. Some of the meteor tracks were out the field of view of the cameras for some of the time and these have also been eliminated from the analysis as either the start or end point of the trail could not be accurately determined.

Complete and reliable data were obtained on four Ursid meteors. The orbital elements, with errors, are obtained for these four meteors. In Table 4 the absolute visual magnitude ($M_V$), the beginning and the terminal heights ($H_b$, $H_e$), the geocentric radiant ($\alpha_g$, $\delta_g$) expressed in J2000.0, and the pre-atmospheric, geocentric, and heliocentric velocities ($V_\infty$, $V_g$, and $V_h$) of these Ursid meteors are given. The SMPN nomenclature of each meteor indicates the date of its detection and its order of detection (i.e. *A* for the earliest and *C* for the latest).

| Meteor Code | $M_v$ | $H_b$ (km) | $H_e$ (km) | $\alpha_g$ (°) | $\delta_g$ (°) | $V_\infty$ (km s$^{-1}$) | $V_g$ (km s$^{-1}$) | $V_h$ (km s$^{-1}$) |
|---|---|---|---|---|---|---|---|---|
| **SPMN 221214A** | -3 | 93.5 | 85.5 | 225.43±0.23 | 74.47±0.21 | 33.8±0.4 | 32.1 | 39.1 |
| **SPMN 231214A** | -4 | 88.6 | 85.6 | 225.4±0.3 | 75.88±0.21 | 32.2±0.4 | 30.1 | 39.0 |
| **SPMN 231214B** | -5 | 96.3 | 69.7 | 229.6±0.5 | 75.9±0.4 | 33.7±0.3 | 31.7 | 40.4 |
| **SPMN 231214C** | -4 | 93.3 | 75.8 | 227.3±0.4 | 76.7±0.4 | 32.2±0.4 | 30.1 | 39.4 |

**Table 4.** Trajectory and radiant data for 4 multi-station Ursid meteors registered in December 2014 (J2000.0).

From these measured radiant positions and velocities, the orbital elements can be derived in a standard way (described in Trigo-Rodriguez et al. (2009), for example). These are given in Table 5. Fig.3 plots the resulting orbit of the meteoroid SPMN 221214A and that of 8P/Tuttle for comparison; together with the SPMN 221214A atmospheric trajectory and its projection on the ground.

| Meteor Code | Day | q (AU) | a (AU) | e | i (º) | ω (º) | Ω (º) |
|---|---|---|---|---|---|---|---|
| **SPMN 221214A** | 22.96484722 | 0.94861±0.001 | 4.16±0.17 | 0.772±0.025 | 51.6±0.5 | 203.3±0.3 | 270.8240 |
| **SPMN 231214A** | 23.03800579 | 0.94500±0.00052 | 3.15±0.24 | 0.700±0.023 | 48.9±0.5 | 205.2±0.3 | 270.8986 |
| **SPMN 231214B** | 23.05556839 | 0.9503±0.0011 | 5.1±0.6 | 0.815±0.021 | 50.3±0.4 | 202.4±0.3 | 270.9165 |
| **SPMN 231214C** | 23.08608796 | 0.9448±0.0011 | 3.5±0.3 | 0.731±0.020 | 48.4±0.4 | 205.0±0.3 | 270.9477 |

**Table 5.** Orbital elements for 4 multi-station Ursids meteors registered in December 2014 (J2000.0).

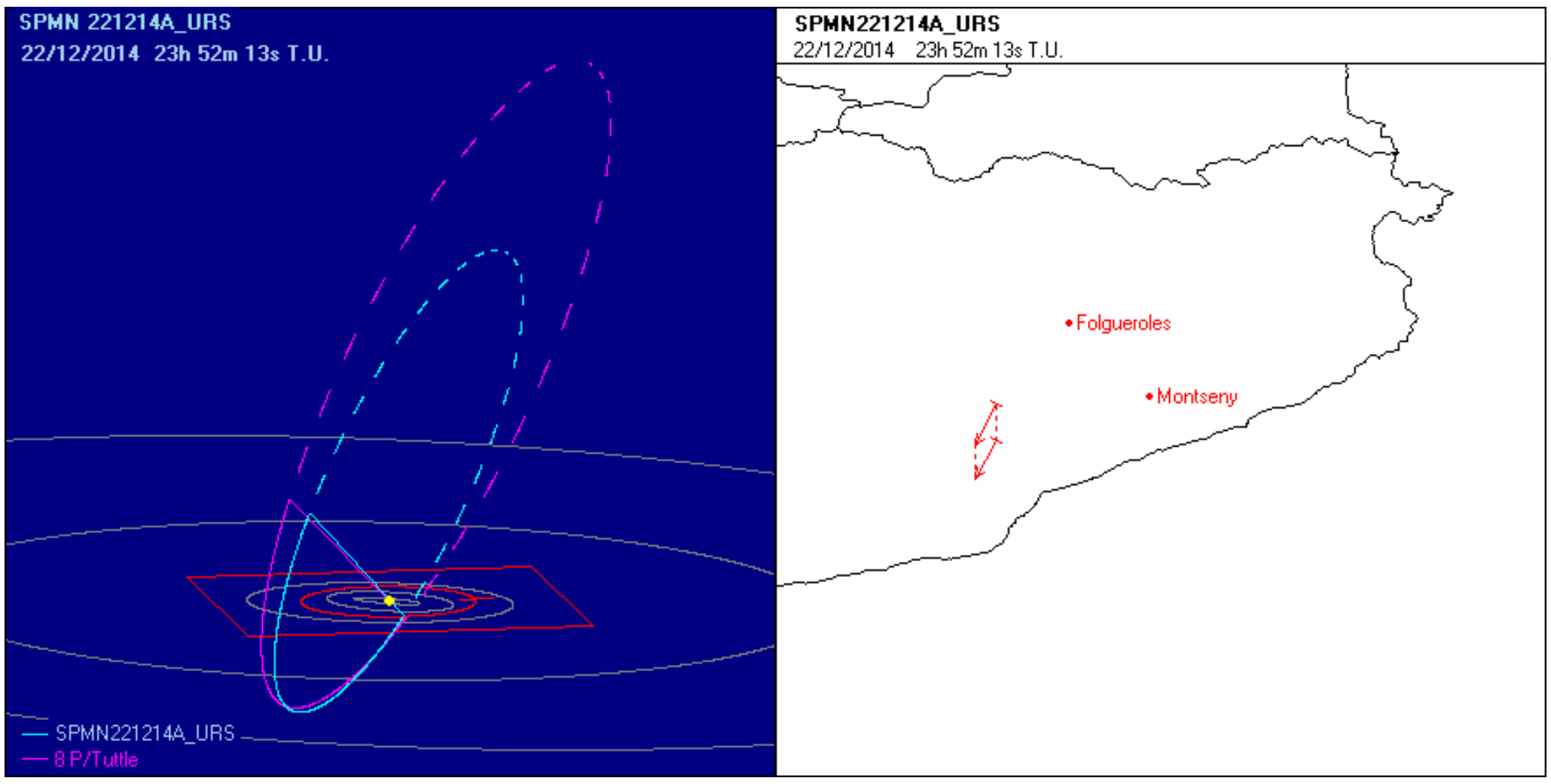

**Figure 3.** (Left) Heliocentric orbit of meteoroid SPMN221214A and comet 8P/Tuttle. (Right) Atmospheric trajectory and its projection on the ground. The SPMN stations from which this meteor was detected are indicated too.

As mentioned in section 2, a simulation of the encounter with possible dust trails was carried out using the software developed by Vaubaillon, Colas and Jorda (2005a, b). A theoretical apparent radiant for the expected encounter was estimated using the Neslušan, Svoreń, and Porubčan (1998) code, method Q (Hasegawa 1990). Table 6 shows a comparison between this value, the average value derived for our four meteors reported, and the previous single-station results of Gajdoš, Tóth and Kornoš (2015). As can be seen, there is excellent agreement between the observational results in this work and the predictions. There is a small discrepancy regarding the right ascension derived by Gajdoš, Tóth and Kornoš (2015), probably explained by the limitations in accuracy of single-station observations. These results can also be compared to previous Ursid outbursts' apparent radiants in 1997 (filament) and 2000 (aphelion outburst) described in Jenniskens et al. (2002). As it can be seen in Fig. 4, the agreement between the value of the averaged apparent radiant presented in this paper and the individual meteor apparent radiants of previous observations is clear.

| **Number of values averaged** | **RA (º)** | **DEC (º)** | **$V_g$ (km $s^{-1}$)** | **Source** |
|---|---|---|---|---|
| Prediction N=19 | 220.85 | +75.4 | 33.21 | J. Vaubaillon (IMCCE) |
| Double station N=4 | 219.85±0.2 | 76.0±0.2 | 32.3 | This work |
| Single station N=19 | 217.9 | +76.4 | - | Gajdoš, Tóth and Kornoš (2015) |

**Table 6.** Predicted apparent radiant positions and averaged geocentric velocity ($V_g$) of the 2014 Ursid dust trail members according to J. Vaubaillon, compared with our double station results and single station results from Gajdoš, Tóth and Kornoš (2015). Equinox (2000.0).

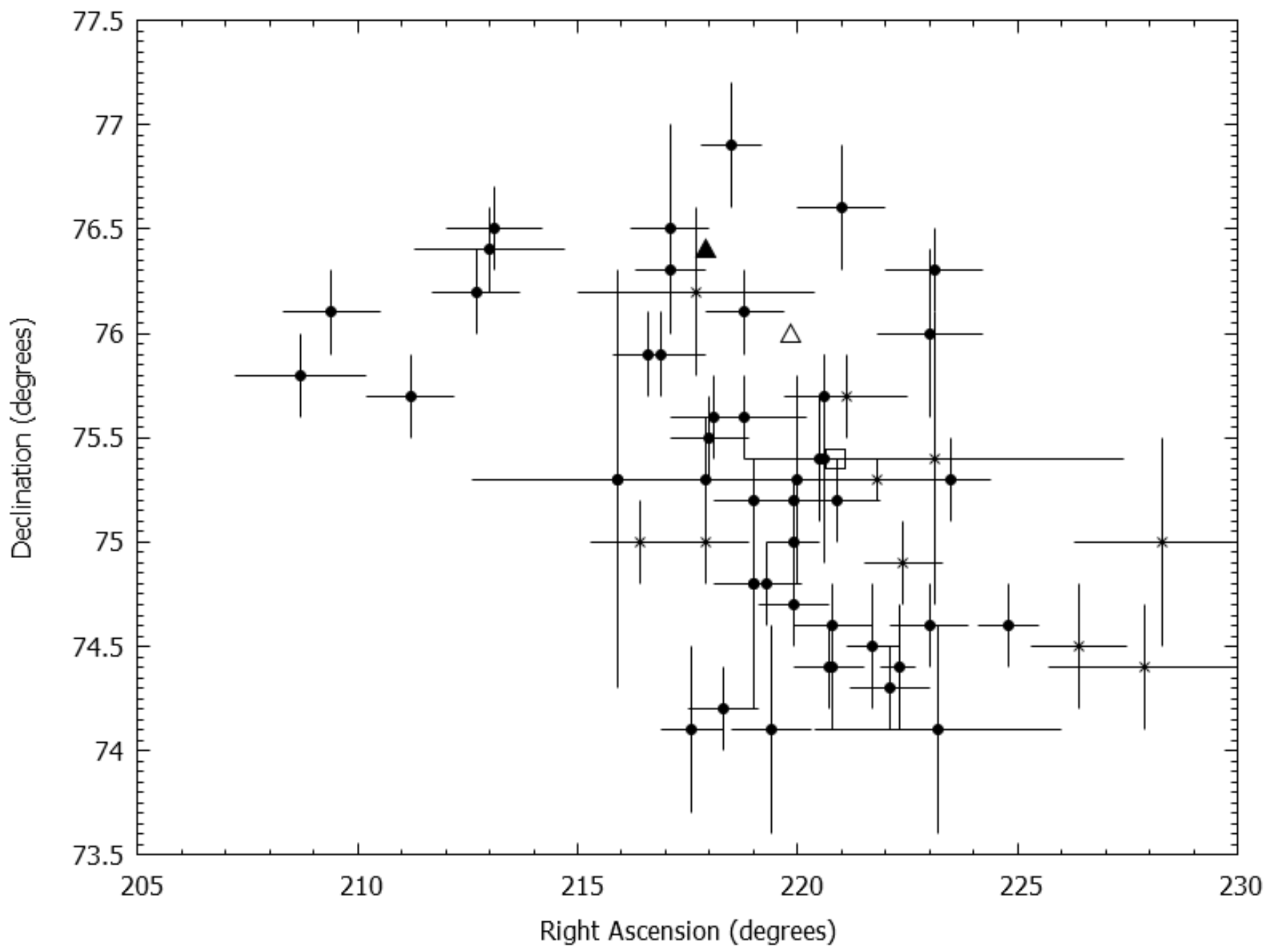

**Figure 4.** The empty triangle marks the averaged apparent radiant derived from our observations; the empty square is the predicted apparent radiant according to J. Vaubaillon software; the filled triangle is the apparent radiant described in Gajdoš, Tóth and Kornoš (2015). The rest of points follow those given by Jenniskens et al. (2002), filled circles mark radiant positions for the 2000 aphelion outburst meteors, and asterisks indicate radiants for the 1997 perihelion outburst meteors.

## 5.- DISCUSSION

Table 7 gives the trajectory including mean beginning and ending height values and orbital elements for the 2000 and the 1997 outburst given by Jenniskens et al. (2002) as well as our results for the 2014 outburst. It can be seen that the SPMN meteor mean trajectory and orbital values slightly differ from those measured in the 2000 outburst.

Fig. 5 is a plot of reciprocal semi-major axis versus perihelion distances for outburst taken from Jenniskens et al. (2002) as well as the four SPMN meteors. As it can be seen, the values are well within the range of the larger sample of past outbursts. Also shown is the location of the 6:7 mean motion resonance. According to Jenniskens et al. (2002), the dust trails producing the outburst when the comet is in its aphelion might be trapped in a 6:7 resonance with Jupiter. The orbits of the four meteors described in this work are slightly closer to the mean motion resonance location than many of the meteors observed in 2000. Two of them (SPMN 221214A and SPMN 231214B) are in fact very close to the resonance line and the other two slightly further away.

Fig. 6 is a plot of inclination against perihelion distance for the same set of meteors as in Fig. 5. Two of the meteors (SPMN 221214A and SPMN 231214B) studied here, have inclinations that are very much in line with the other observed meteors while two (SPMN 231214A and SPMN 23124C) have inclinations below 49$^{\circ}$, well below any other recorded value for the Ursids.

| Year | 1997 (10 orbits) Filament | 2000 (59 orbits) Aph. outburst | 2014 (4 orbits) Aph. Outburst |
|---|---|---|---|
| **Date** | Dec. 22.434 | Dec. 22.32 | Dec. 23.0375 |
| **$H_b$** | 104.9 | 107.1 (52 orbits) | 92.93 |
| **$H_e$** | 94.2 | 96.2 (52 orbits) | 79.15 |
| **RAgeo** | 222.1 | 219.0 | 219.85 |
| **DECgeo** | 75.0 | 75.3 | 76.0 |
| **Vgeo** | 32.25 | 33.05 | 32.3 |
| **a** | 4.62 | 4.673 | 3.978 |
| **e** | 0.795 | 0.799 | 0.755 |
| **q** | 0.944 | 0.940 | 0.956 |
| **i** | 51.5 | 52.5 | 49.8 |
| **ω** | 204.9 | 205.9 | 203.9 |
| **Ω** | 270.64 | 270.76 | 270.9 |

**Table 7.** Averaged date, beginning and ending heights, geocentric radiant position, geocentric entry velocity and orbital elements for the filament observations in 1997 and the aphelion outburst observation in 2000 (Jenniskens et al. (2002)) and 2014 (presented in this work).

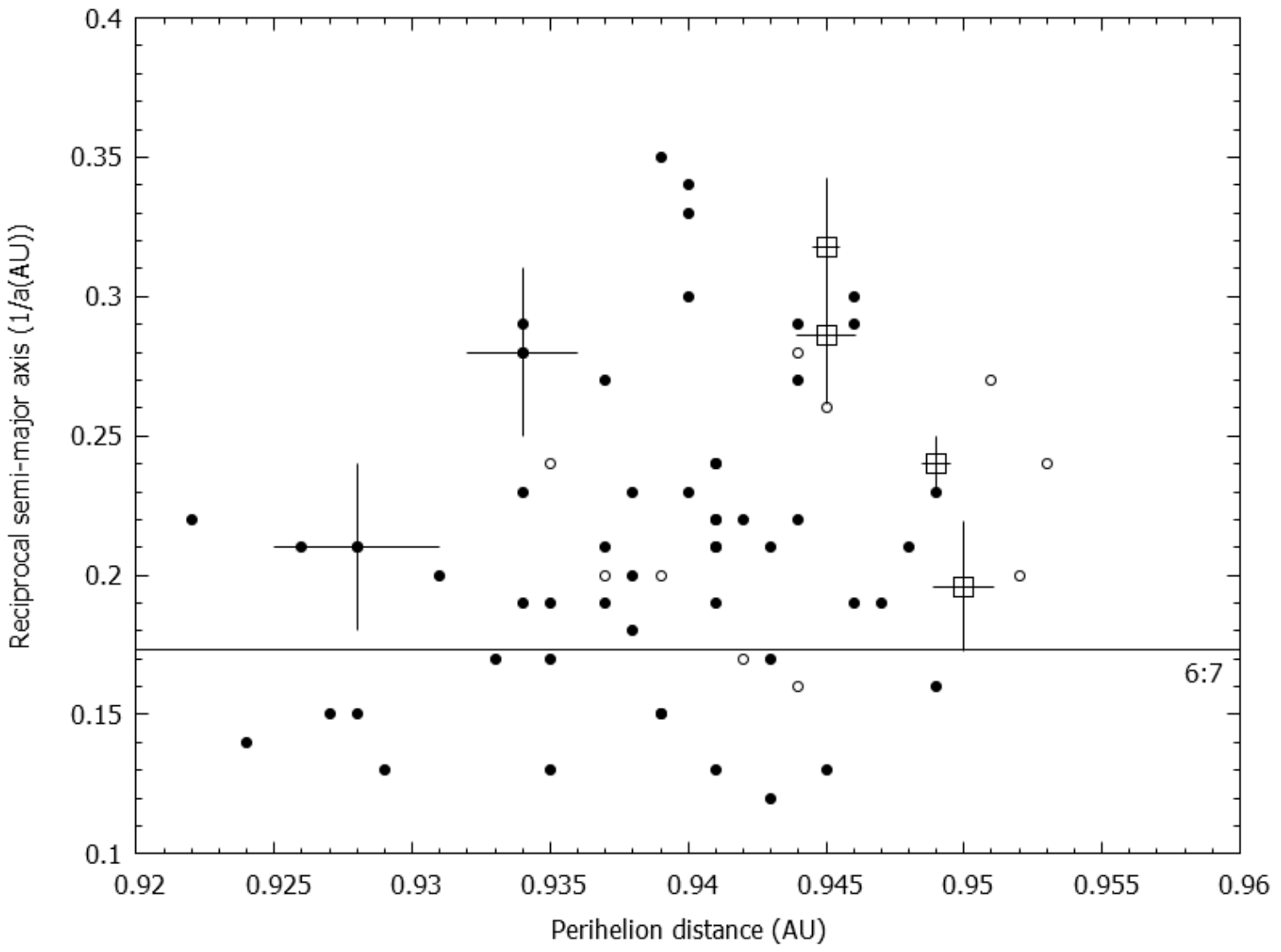

**Fig. 5.** A plot of the perihelion distances versus the reciprocal semi-major axis for the meteors registered at: the Ursid aphelion outburst in the year 2000 (filled circles) as in Jenniskens et al. (2002); the perihelion outburst of 1997 (empty circles) as in Jenniskens et al. (2002); and the four Ursids meteors studied in this work (open squares). Error bars of the four SPMN meteors and a couple of representative error bars of 2000 outburst data reported in Jenniskens et al. (2002) are plotted. The straight line indicates the resonance 6:7 inverse semi-major axis.

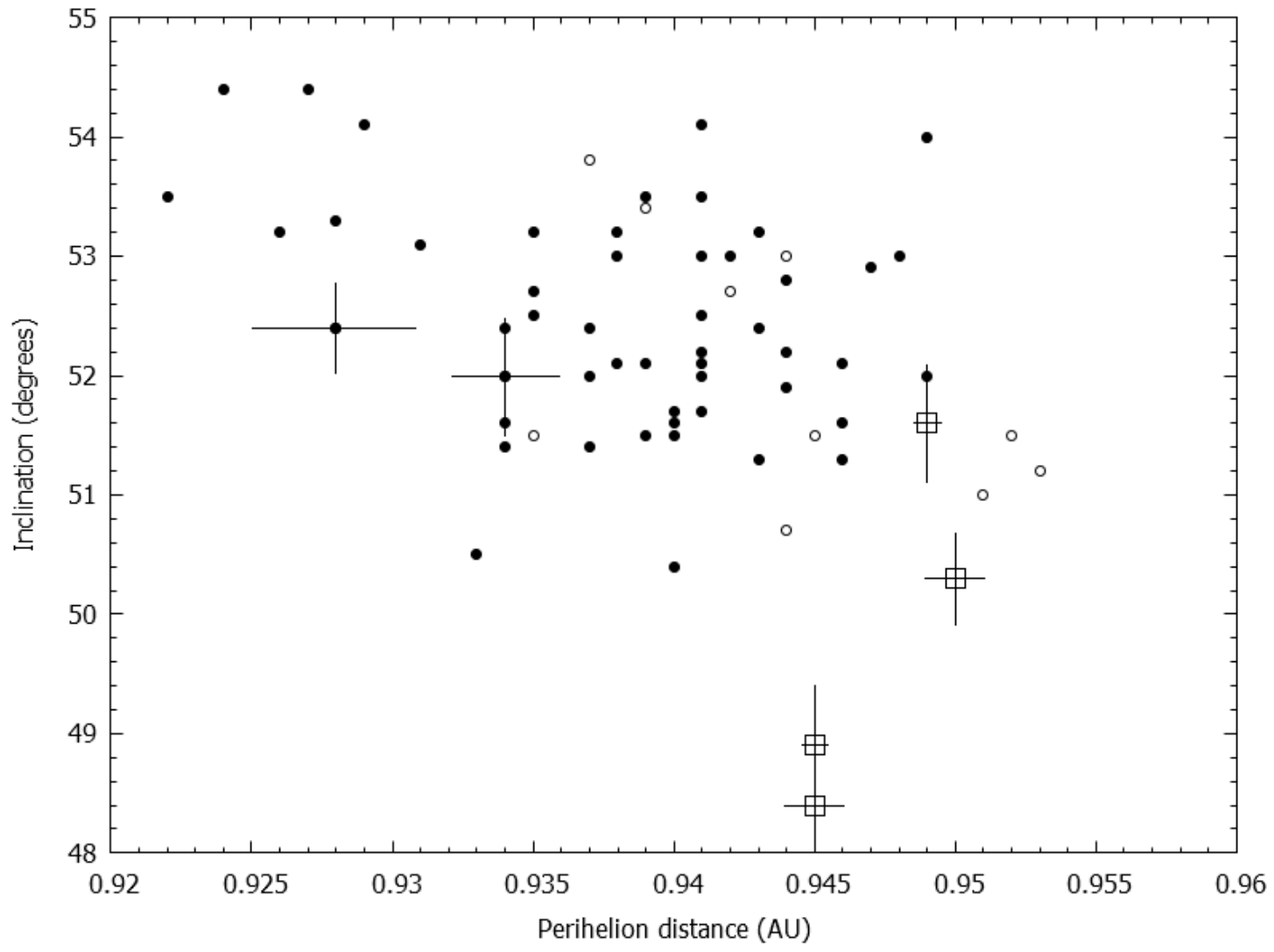

**Fig. 6.** A plot of the perihelion distances versus orbit inclination for the meteors registered at: the Ursid aphelion outburst in the year 2000 (filled circles) as in Jenniskens et al. (2002); the perihelion outburst of 1997 (empty circles) as in Jenniskens et al. (2002); and the four Ursids meteors studied in this work (open squares). Error bars of the four SPMN meteors and a couple of representative error bars of 2000 outburst data reported in Jenniskens et al. (2002) are plotted.

It is difficult to judge whether the differences are significant or not, given the small size of our sample in 2014. We have two meteors (SPMN 221214A and SPMN 231214B) that are very much in agreement with all the other evidence and two that do not fit so well. With a much larger sample it would be easy to determine whether the two discrepant ones are outsiders or not.

The discrepancies found between some of the observed trajectory and orbital data and the large sample of Jenniskens et al. (2002) arise because of the slow geocentric velocity of these SPMN meteoroids. Meteor detection and measurement depends on the meteor luminosity, which is, in general, proportional to the body pre-entry mass, the initial velocity to the power of three and the sine of the slope between the horizon and the trajectory (Gritsevich and Koschny 2011; Bouquet et al. 2014). Thus, the slow geocentric velocity of some Ursid meteors could have affected the beginning velocities and heights assigned to them (they did not become visible to detectors until they had penetrated further into the atmosphere). The orbital semi-major axis, the eccentricity and the inclination are derived from the entry velocity and the trajectory slope. This could particularly be the case of SPMN 231214A and SPMN 231214C which we have already highlighted as being potentially discrepant.

If the values we have derived for all four meteors are correct, then we should consider the possible effect of close Earth encounters on the orbits of the dust trails. The dust trails are closest to Earth when they are near perihelion and will thus be travelling with a slightly higher heliocentric velocity than the Earth. The cumulative effect of repeated encounter will be to decelerate the meteoroid, (for a mathematical formulation of the effects of changing orbital energy on meteoroids, see Williams (2002, 2004)). This could explain the discrepancy. Unfortunately, the lack of reliable data on orbits of Ursid meteor aphelion outburst before AD 2000 combined with this small meteor sample, makes it impossible to confirm that this occurred in the Ursids.

Finally, it could be argued some of the four meteors reported here (especially the two discrepant ones) belong to a different cometary material detachment. An encounter with the filament material was predicted to occur in practically the same date as the aphelion outburst. Despite the broader shape of the filament, previous studies (Jenniskens, 2006) indicate their higher inclination orbits, which is the opposite of what was found.

## 6.- CONCLUSIONS

We have presented accurate orbital information based on double-station detections of four Ursid meteors detected during the last Ursid outburst predicted for December 2014. This outburst corresponds to cometary material detached from the comet 8P/Tuttle in one of its perihelion transits that encounters the Earth when the comet is close to its aphelion. The results can be summarized as following:

- Video and forward-scatter detections along with other reports indicate high meteor activity associated with an Ursid dust trail crossing the Earth's orbit at solar longitude at $\lambda_0$=271.8º on Dec. 23rd, 2014.
- The outburst was characterized by relatively large meteoroids (population index of 1.8), producing bright meteors and some fireballs that were recorded by our all-sky systems and video cameras.
- The mean Ursid ZHR was around 19 meteors/hour, peaking with a ZHR of 45 ±19 at around solar longitude $\lambda_0$=271.85º.
- Two of the four Ursid orbits (SPMN 221214A and SPMN 231214B) exhibit similar orbital elements to the previously recorded meteoroids during outbursts. The other two were measured slightly below the expected geocentric velocity, and their measurements were probably affected by low meteor brightness on the very beginning trajectory segment. However, the meteor orbits retrieved are well within the range of values of previous aphelion outbursts.
- Despite some minor inaccuracies, the four meteoroids have orbits that seems to be associated with the 1405 or 1392 dust trails which provoked the outburst and are captured in the two body mean motion resonance with Jupiter.

**Acknowledgments**

This study was supported, by the Spanish grants AYA2011-26522 and AYA2015-67175-P (PI: JMTR), AYA2015-68646-P and AYA2014-61357-EXP (JMM). The FRIPON project is funded by ANR. MG acknowledges support from the ERC Advanced Grant No. 320773, and the Russian Foundation for Basic Research, project nos. 16-07-01072 and 16-05-00004. Research at the Ural Federal University is supported by the Act 211 of the Government of the Russian Federation, agreement No 02.A03.21.0006. EB is supported by the Universitat Ramon Llull project 2016-URL-IR-001 supported by "Generalitat de Catalunya". The authors acknowledge being a part of the network supported by the COST Action TD1403 "Big Data Era in Sky and Earth Observation". The CINES supercomputer was used for theoretical work. We thank Dr. Aswin Sekhar for his valuable comments that helped to enhance this paper. This study was done in the frame of a PhD. on Physics at the Autonomous University of Barcelona (UAB) under the direction of Dr. Josep Ma. Trigo-Rodríguez and Dr. Maria Gritsevich.